\documentclass[aps,prb,floats,amsmath,showpacs,twocolumn]{revtex4}
\usepackage{graphicx}  
\begin{document}

\title{Theoretical investigation of hydrogen storage in
metal-intercalated graphitic materials}

\author{Manuel Cobian and Jorge \'I\~niguez}

\address{Institut de Ci\`encia de Materials de Barcelona (ICMAB-CSIC),
Campus UAB, 08193 Bellaterra, Spain}

\begin{abstract}
We have used first-principles methods to investigate how metal atoms
dispersed in the interlayer space of graphitic materials affect their
hydrogen-binding properties. We have considered ideal stage-one
metal-intercalated graphites of various compositions as representative
model systems. Our calculations suggest that alkaline earth metals can
significantly enhance the hydrogen storage properties: for example, Be
and Mg atoms would act as binding sites of three or four hydrogen
molecules, with binding energies per H$_2$ in the 0.2--0.7~eV range,
as required for applications. We also find that alkali and transition
metals are not as effective in enhancing the storage capacity.
\end{abstract}


% 84.60.-h   energy storage
% 81.05.Uw   graphite
% 71.20.Tx   intercalation compounds
% 71.15.Pd   Molecular dynamics calculations, condensed matter

\maketitle

\section{Introduction}

Developing practical means to store hydrogen at ambient conditions is
critical for the progress of fuel-cell technologies. The possibility
of storing hydrogen in carbons attracts great interest, as they would
constitute a light and cheap storage medium susceptible of large scale
production. Unfortunately, the dominating carbon--hydrogen
interactions that occur in the usual materials are not suitable for
storage purposes: H$_2$ physisorption is too weak to retain the
hydrogen at ambient conditions, and chemisorption of H atoms is too
strong and hampers the hydrogen release.\cite{hir03} Current efforts
thus focus on obtaining binding interactions in the range that would
be suitable for applications, namely, between 0.2 and 0.8~eV.

Recent {\sl ab initio} simulations have suggested that
carbon-supported transition\cite{yil05a,zhao05,yil05b} and
alkali\cite{sun06} metal atoms bind hydrogen with the appropriate
interaction energy (e.g., about 0.5~eV per H$_2$ in Ti-decorated
nanotubes) and are acceptable from a gravimetric viewpoint. Moreover,
it seems that the predicted binding mechanism has been observed
experimentally in multiwalled nanotubes on which nickel nanoparticles
were dispersed.\cite{kim05,lee06} Those promising results clearly
indicate this storage strategy deserves further study. However, it is
also evident that the new metal-mediated hydrogen-storage mechanisms
will be useful only if they are present in materials that are easy to
produce (e.g., the usual activated carbons or graphitic nanofibers)
and free from problems that could impede a practical application
(e.g., oxidation or segregation\cite{sun06} of the metal atoms). So
far, the simulations have focused on cases in which the metal atoms
are deposited on free-standing nanotubes or fullerenes. Such model
systems may be expected to mimic the situation in the walls of the
large pores present in many carbons. Unfortunately, the calculations
indicate that, except for the case of Li, such materials are bound to
suffer from metal segregation and clustering problems.

A possible strategy to diminish the segregation problem is to look for
materials in which the metal atoms are more tightly bound to the
carbon structure. That should be the case of metal-intercalated
graphites, graphitic nano-fibers, and carbons of the type used for
electrodes in lithium batteries. Continuous progress in the synthesis
of intercalated graphites (which is best illustrated by the recent
obtention\cite{emery05a,emery05b} of massive samples of stage-one
crystalline C$_6$Ca) and carbons for battery electrodes provides
encouragement for a theoretical investigation of such possibilities,
as it suggests that predictions of new materials may be experimentally
realizable.

We have thus studied this general class of materials, in which the
metal atoms can appear {\sl sandwiched} between graphene layers, using
first-principles simulation methods. Since this work aims at
identifying trends as a function of the metal species and quantifying
the dominant interactions involving hydrogen, we have focused on the
simplest model systems capturing the relevant chemical and physical
effects: {\sl stage-one} metal-intercalated graphites of various
compositions. (The term {\sl stage-one} implies that all the
interlayer spaces have the same metal content.) Note that, strictly
speaking, some of the graphites considered might be impossible to
obtain experimentally because of staging effects\cite{dresselhaus81}
that we do not intend to address here. Yet, the phenomena occurring in
the vicinity of the metal atoms in our simulated systems can be
expected to occur also in the other carbons mentioned above.

The paper is organized as follows. Section~II describes the employed
methodology, with special emphasis on the status of Density Functional
Theory\cite{hohenberg64,kohn65} (DFT) methods regarding van der Waals
interactions and the practical approach adopted here. In Section III
we present and discuss our results for alkali (Subsection A), alkaline
earth (Subsection B), and representative transition (Subsection C)
metals. In Subsection~D we discuss results pertaining to the binding
of the metal atoms to the carbon structure and the possibility that
segregation occurs. Finally, in Section~IV we give our summary and
conclusions.

\section{Methodology}

We used the Local Density (LDA) and Generalized Gradient (GGA)
Approximations to Density Functional Theory as implemented in the code
VASP.\cite{vasp} More precisely, we used the Perdew-Zunger
parametrization\cite{perdew81} of Ceperley-Adler
data\cite{ceperley80} for the LDA, and the so-called GGA-PBE of
Ref.~\onlinecite{perdew96}. The PAW scheme\cite{paw1,paw2} was
employed to represent the atomic cores. Nominal valence electrons were
explicitly treated, as well as the semi-core electrons of all the
metal atoms considered (e.g., the 3$s$ and 3$p$ electrons of Sc). We
used a plane-wave basis with a 400~eV cut-off, and k-point grids with
spacings that were in all cases smaller than 0.07~\AA$^{-1}$; we
checked these calculation conditions lead to sufficiently converged
results. We also used the first-principles software package
PWscf\cite{pwscf} to analize the electronic properties of the most
relevant structures obtained with VASP. We found spin polarization
only in a few of the systems considered (the metal-doped single
graphene layers and the graphites intercalated with n$d^3$ transition
metals). Thus, the vast majority of the studied materials, including
the most relevant ones for hydrogen storage purposes, contain no
unpaired electrons.

The van der Waals interactions, responsible for the physisorption of
H$_2$ in carbons, are likely to play an important role in many of the
systems considered here. The status of DFT in relation to van der
Waals forces is an awkward one: while it is known that the usual LDA
and GGA's are ill-suited to model such interactions, for many
physisorption systems the LDA renders binding energies that are
(albeit accidentally) in reasonable quantitative agreement with the
experiment.\cite{arellano00} (The GGA, on the other hand, usually
renders significantly smaller or even repulsive
interactions.\cite{tada01,henwood07}) More specifically,
Ferre-Vilaplana\cite{vilaplana05} has performed careful (and
computationally costly) MP2 studies correcting for basis-set
superposition errors, and found that the binding energy of H$_2$
physisorbed on graphene is about 0.064~eV, to be compared with the LDA
result of 0.083~eV.\cite{arellano00} While this result clearly
indicates a serious overbinding associated to the LDA, it also
suggests the LDA binding energies can be considered as qualitatively
correct, with a {\sl systematic} quantitative error of about
30\%. Having this in mind, many authors have regarded LDA as a
reasonably reliable and computationally efficient method to study this
type of systems. Here we adopt the same approach. Finally, note that
for some of the materials studied -- most importantly, for the
graphites intercalated with alkaline-earth metals --, we found that
the dominant interactions are not of a dispersive nature. In such
cases, in which DFT can be expected to be quantitatively accurate, we
performed GGA calculations to ratify the LDA results.

It should also be noted that in our simulations we did not considered
the quantum mechanical nature (i.e., quantum-rotor effects) of the
hydrogen molecules. Such a simplified representation becomes accurate
whenever the molecules are tightly bound or the interactions are as
large as to cause H$_2$ dissociation. For all other cases, the
approximation of treating the hydrogen molecules as classical objects
must be born in mind. Note also that we did not consider thermal
effects in the calculations and, thus, all the obtained results
correspond to the zero temperature limit and directly reflect the
fundamental interactions.

Our simulations were aimed at the computation of the binding energy
between hydrogen molecules and various metal-intercalated
graphites. Binding energies ($E_b$'s) are computed as the energy
difference between the bound system and its constituents. For example,
the binding energy of H$_2$ to the C$_4$Li intercalated carbon is
given by $E$(H$_2$) $+$ $E$(C$_4$Li) $-$ $E$(C$_4$Li-H$_2$), where
$E$(H$_2$) is the energy of the isolated hydrogen molecule relaxed to
its equilibrium (lowest-energy) configuration, etc. The H$_2$
molecules were introduced in the system one at a time, which allowed
us to compute the binding energy as a function of hydrogen load.

\begin{figure}[top!]
\includegraphics[width=\columnwidth]{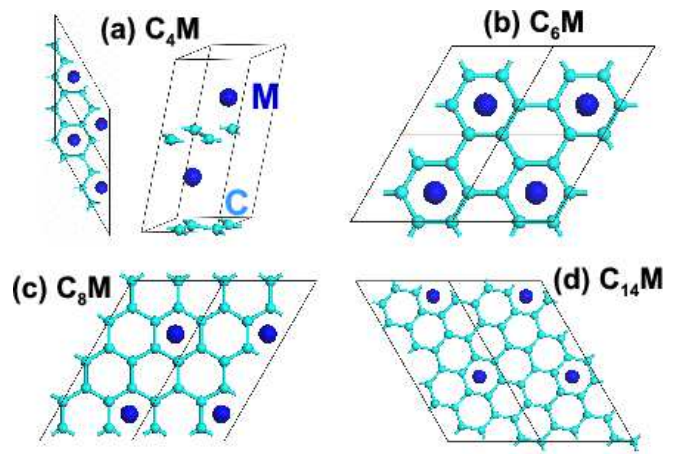}
\caption{(Color online) Schematic representation of the four basic
  model systems considered in this study. A top view of a 2$\times$2
  in-plane repetition of the supercell is shown in all the cases; for
  the C$_4$M composition in panel~(a), a lateral view of the supercell
  is also shown.}
\label{fig1}
\end{figure}

Figure~\ref{fig1} represents the main model supercells considered,
which correspond, respectively, to the chemical formulas C$_4$M,
C$_6$M, C$_8$M, and C$_{14}$M, where M is the metal atom. Note that,
as shown in panel~(a) for the C$_4$M case, our supercells included two
graphene layers (and, thus, two interlayer spaces with metal atoms)
whose relative position was allowed to evolve freely towards the
lowest-energy solution during the structural relaxation. Given the
intricate energy landscape associated to the materials considered, we
found it necessary to perform the structural relaxations in a very
meticulous way: We first ran a short (50~fs) constant-energy molecular
dynamics with initial random velocities corresponding to a temperature
in the 100-700~K range; the obtained structure was then used as the
starting point of a full relaxation of atomic positions and cell
parameters. The structural relaxations were regarded as completed when
residual forces (stresses) were smaller than 0.05~eV/\AA\ (0.05~GPa);
we checked this criterion leads to sufficiently accurate energies. For
all the systems considered, the described procedure was repeated a
minimum of 7 times (up to a maximum of 40 times in specially difficult
or significant cases) starting from different initial configurations,
from which only the lowest-energy solution was taken into account for
further considerations. (We wanted to study the behavior of these
materials at room and higher temperatures; thus, restricting ourselves
to the the lowest-energy solutions seemed perfectly justified.) The
initial configurations were carefully chosen so as to cover a range of
possibilities as broad as possible; in particular, for all the
compositions studied, we considered initial configurations involving
molecular and atomic hydrogen, so as to be able to determine in which
cases H$_2$ dissociation is energetically favorable. We explicitly
checked this relaxation scheme leads to the correct result in {\sl
simple} cases with only one H$_2$ molecule {\sl per} metal atom, for
which it is possible to identify the global minimum by starting
relaxations from all possible critical points (i.e., maxima, minima
and saddle points) of the complex energy landscape of the system.

It is worth to comment on the possible errors associated to our
particular choice of model systems and relaxation scheme. The size of
the considered supercells determines the number of independent atoms
in our simulations, defining the configuration space explored in the
structural relaxations. Having relatively small supercells may thus be
problematic, specially in cases in which H$_2$ molecules dissociate
and form metal--hydrogen complexes. Being aware of this, we checked
the suitability of the smallest supercell considered (shown in
Fig.~\ref{fig1}a) by repeating the study of some compositions using a
supercell twice as big (doubled in the plane). We did not find any
qualitative changes with respect to the results obtained with the
small supercell, which suggests we can trust the qualitative
correctness of our results. At any rate, it is important to realize
that these finite-size errors, as well as other possible errors in the
identification of the global minima, inevitably become worse as the
number of H$_2$ molecules increases, implying that our calculated
hydrogen-binding energies will be smaller than the ones we would {\sl
ideally} obtain. Hence, errors of this nature are actually quite
benign, as they result in an underestimation of the H$_2$-storing
ability of these materials.

\section{Results and discussion}

We find that the electronic structure of the metal atom completely
determines the system's hydrogen-storage properties. Accordingly, we
have divided our results in three groups, which correspond to the
alkali, the alkaline earth, and the transition metals, respectively.

\subsection{Alkali metals}

\subsubsection{Simulation results}

\begin{table}[bottom!]
\caption{Calculated binding energies (in eV) of hydrogen molecules
  absorbed in stage-one graphites intercalated with alkali metal
  atoms. We considered the insertion of H$_2$ molecules in the system
  one at a time, so as to compute the dependence of the binding energy
  with the hydrogen load. For example, the result for C$_6$Li-nH$_2$
  and n=2 is obtained from
  $E$(C$_6$Li-H$_2$)~$+$~$E$(H$_2$)~$-$~$E$(C$_6$Li-2H$_2$). For each
  system, we show results up to the maximum number of H$_2$ molecules
  per metal atom that can be absorbed (n$_{\rm max}$), which varies
  with the carbon-metal ratio. The average binding energy per molecule
  ($\bar{E}_b$) is obtained from the fully hydrogen-loaded result as
  [$E$(C$_6$Li)~$+$~n$_{\rm max}$~$E$(H$_2$)~$-$~$E$(C$_6$Li-n$_{\rm
  max}$H$_2$)]/n$_{\rm max}$. We also indicate the hydrogen weight
  percentage corresponding to n=n$_{\rm max}$.}
\label{tab1}
\vskip 2mm

\begin{tabular*}{\columnwidth}{@{\extracolsep{\fill}}lrrrrrc}
\hline\hline
System            & n=1     & n=2    & n=3    & ... n=6     & $\bar{E}_b$  & wt.\% \\
\hline
C$_6$Li-nH$_2$    & $-$0.23 & 0.18   &        &         & $-$0.03             &  4.9        \\
C$_8$Li-nH$_2$    & $-$0.14 & 0.16   & 0.31   &         &    0.11             &  5.6        \\
C$_{14}$Li-nH$_2$ & $-$0.27 & 0.23   & 0.22   & 0.23    &    0.14             &  6.5        \\
\hline
C$_6$Na-nH$_2$    &    0.31 & 0.30   &        &         &    0.30             &  4.1        \\
C$_8$Na-nH$_2$    &    0.29 & 0.32   & 0.31   &         &    0.31             &  4.8        \\
C$_{14}$Na-nH$_2$ &    0.24 & 0.29   & 0.33   & 0.26    &    0.27             &  6.0        \\
\hline
C$_6$K-nH$_2$     &    0.32 & 0.24   &        &         &    0.28             &  3.5        \\
C$_8$K-nH$_2$     &    0.32 & 0.34   & 0.17   &         &    0.28             &  4.3        \\
C$_{14}$K-nH$_2$  &    0.28 & 0.38   & 0.28   & 0.21    &    0.26             &  5.5        \\
\hline\hline
\end{tabular*}
\end{table}

\begin{figure}[top!]
\includegraphics[width=\columnwidth]{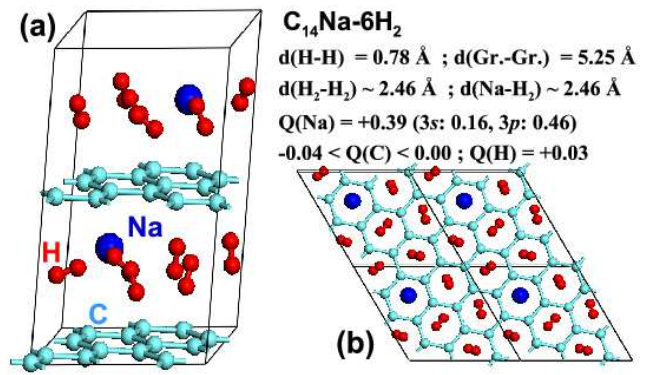}
\caption{(Color online) Lateral (a) and top (b) views of the
  lowest-energy configuration computed for the C$_{14}$Na-6H$_2$
  composition. In panel~(b) a 2$\times$2 in-plane repetition of the
  considered supercell is shown. The relevant distances and L\"owdin
  charges (in elementary charge units) are given. The in-plane lattice
  parameters are essentially unaffected by the insertion of metal and
  hydrogen atoms.}
\label{fig2}
\end{figure}

Table~\ref{tab1} summarizes our results for the binding energies and
Fig.~\ref{fig2} shows the lowest-energy configuration of
C$_{14}$Na-6H$_2$, which is a representative case. In the Na
compounds, we find that the binding energy of all the absorbed
molecules is about 0.3~eV and is largely independent of the amount of
hydrogen present. The situation is similar for K, although the
dependence of $E_b$ with the amount of stored hydrogen is more
significant. In the case of Li, on the other hand, the binding energy
of the first H$_2$ molecule is negative, implying that the absorption
is not energetically favorable. However, the absorption of additional
molecules is predicted to occur, with binding energies of as much as
0.3~eV. (This seemingly puzzling behavior is explained in the {\sl
Interpretation} section below.) Finally, the metal content seems to
have a relatively small influence in the binding properties, as we
obseve no drastic effect associated to the C--M ratio.

The obtained lowest-energy solutions are, without exception,
characterized by the following two features: (1) The graphene layers
are perfectly superimposed in an {\sl eclipsed} configuration (i.e.
an ``AA stacking''). (2) The hydrogen molecules position themselves in
the space between two eclipsed carbon hexagons, and tend to align with
the direction perpendicular to the graphenes. The observed deviations
from an homogeneous alignment of the H$_2$ molecules are probably the
reflection of a complex multi-minima energy landscape, determined by
small H$_2$--H$_2$ and H$_2$--carbon/metal interactions whose
magnitude falls below the accuracy requested in our structural
relaxations. As indicated in the previous Section, we explicitly
considered the possibility of H$_2$ dissociation by chosing
appropriate initial configurations for the structural relaxations
(i.e. configurations in which we start with two separated H atoms
instead of an H$_2$ molecule). However, from such calculations we
always obtained either a higher-energy solution or a recombination of
the H atoms to form a molecule. Moreover, we never observed H$_2$
dissociation in relaxations where the starting point was molecular
hydrogen.

In what regards the electronic structure of the lowest-energy
configurations, a standard analysis in terms of L\"owdin
charges\cite{lowdin50} indicates that the cation donates most of its
valence electron to the graphene layers, which are thus negatively
charged. Interestingly, the charges computed for the metal and carbon
atoms remain essentially unaffected by the insertion of H$_2$
molecules; accordingly, the molecules do not charge significantly, and
an analysis of the electronic wave functions indicates very weak
bonding with the carbon or metal atoms. (Such weak bonds are typical
of LDA simulations of van der Waals systems. While known to be an
artifact of the theory, they render an interaction of a magnitude that
is similar to that of the actual van der Waals forces, as mentioned in
Section~II.) The calculated H--H distances are always around 0.78~\AA,
closely resembling the 0.77~\AA\ that the LDA renders as the
equilibrium bond distance of an isolated H$_2$ molecule.

\subsubsection{Interpretation}

To understand these results, it is useful to consider the absorption
of molecular hydrogen in graphite when no metal atoms are present. In
that case, our LDA scheme predicts a binding energy of 0.13~eV per
molecule in the fully-loaded situation corresponding to a CH chemical
formula. (We denote as ``fully loaded'' the case in which there is one
H$_2$ molecule in every {\sl site} between C$_6$ hexagons.) On the
other hand, when only a few H$_2$ molecules are inserted into the
structure, the graphite--H$_2$ interaction turns out to be repulsive
(we computed a negative binding energy of $-$0.07~eV per molecule for
the C$_8$H$_2$ system). These results, which agree with earlier
theoretical studies,\cite{jacobson02} reflect the competition of
interactions that is at play in the hydrogen sorption process. In
essence, the insertion of hydrogen causes an increment of the
graphene--graphene interlayer distance, which switches from the
3.4~\AA\ computed for pure graphite to the 5.2~\AA\ obtained for the
fully hydrogen-loaded material. The absorption of only a few H$_2$
molecules is thus energetically unfavorable, as the energy gain
associated to the graphene--H$_2$ interaction cannot compensate for
the energy cost associated to the increased interlayer
distance. (Roughly, our results indicate that the interlayer binding
energy in graphite is about 0.02~eV per C atom, and that the
graphene--H$_2$ interaction energy {\sl within graphite} is about
0.11~eV.) As more hydrogen goes into the system, by applying H$_2$
pressure experimentally, the energy balance switches sign and we
obtain a net binding energy per H$_2$ molecule.

In our alkali metal-intercalated graphites, the interlayer distances
in the systems without hydrogen are 3.4, 4.3, and 4.9~\AA,
respectively, for Li, Na, and K (the values correspond to the
materials with C$_{12}$M chemical formula and are representative for
the rest). Thus, it is not surprising that the materials with Na and K
can uptake H$_2$ molecules, since the interlayer distance is nearly
unaffected by the hydrogen absorption and, thus, there is no energy
cost associated to an interlayer expansion. The case of Li, on the
other hand, is very similar to that of metal-free graphite, i.e., the
absorption is favorable only when a large enough number of hydrogen
molecules are involved. This picture is essentially identical to the
one proposed to explain early experimental results for the absorption
of hydrogen and larger molecules in K-intercalated
graphites.\cite{watanabe73}

The above reasoning suggests it might be possible to store hydrogen
molecules at binding sites that do not have any neighboring metal
atom, provided that the interlayer distance is large enough. We thus
considered systems of composition C$_{18}$M-nH$_2$, in which binding
sites isolated from the metal atoms exist, and performed the kind of
study described above. In the case of Na, the computed binding energy
was 0.27~eV for H$_2$ molecules neighboring a metal atom, and 0.23~eV
for those that do not have any metal as first neighbor. We obtained
the same qualitative results for the Li (0.13~eV average $E_b$ for
molecules neighboring a metal atom and 0.19~eV for the additional
ones) and K (with binding energies of 0.26~eV and 0.22~eV,
respectively) systems. These results strongly support the notion that
the enhancement of the H$_2$ binding is related to the increase of the
interlayer distance caused by the metal atoms, and not to a direct
metal--H$_2$ coupling. It is worth to mention here that this
conclusion contrasts with theoretical studies of Li-doped graphene: in
that case, an {\sl enhanced-physisorption} type of interaction was
found for H$_2$ molecules in the vicinity of the alkali metal atom,
the enhancement being associated to a significant charge transfer to
hydrogen.\cite{cabria05} In addition, electrostatic interactions
between free-standing alkali metal atoms and H$_2$
molecules,\cite{lochan06} associated to quadrupoles and induced
dipoles, are known to exist, and one might have expected them to play
a role in the systems here considered. However, our results indicate
that such charge transfer and electrostatic effects are secondary for
the hydrogen interactions in graphites intercalated with alkali
metals.

We performed a second computer experiment aimed at ratifying the above
conclusions. We considered the system that can be described as the
periodic repetition of the C$_6$--Na--C$_6$-- motif, in which half of
the interlayer spaces are free from metal atoms. (This would
correspond to a {\sl stage-two} intercalation.) In this case, we
obtained an average binding energy of 0.11~eV for the molecules in the
metal-free interlayer space, which is very close to the value reported
above for the absorption of H$_2$ in pure graphite. This result
confirms that it is the expanded interlayer distance, and not the fact
that the graphene layers are charged, what facilitates the binding of
H$_2$ molecules in these systems.

\subsubsection{Discussion and connection with experiment}

Our simulations indicate that alkali metals intercalated in graphite
enhance the binding properties of the system, and that the effect
holds for a wide range of carbon--metal ratios. We find that the metal
atom plays the role of increasing the interlayer distance, thus
allowing a more energetically favorable absorption. As mentioned
above, such a conclusion is essentially in agreement with what was
experimentally observed\cite{watanabe73} in studies of the absorption
properties of K-intercalated graphites.

The experimental work on state-two C$_{24}$K offers some quantitative
data that can be used to check the soundness of our simulations. Most
significantly, it is observed experimentally that, upon absorption of
H$_2$, the inter-graphene distance expands by about
0.29~\AA.\cite{watanabe73} This is compatible with our results for
stage-one C$_{14}$K, the closest composition we have studied, for
which we observe that the addition of two H$_2$ molecules per metal
atom shifts the interlayer spacing from 4.90~\AA\ to 5.17~\AA. The
measured heat of absorption, on the other hand, is about 2.3~kcal/mol
(i.e., 0.10~eV per H$_2$ molecule).\cite{watanabe73,cheng01} Such a
quantity is not directly comparable to the binding energies of about
0.28~eV per H$_2$ that we obtained for the compounds with K, since we
did not include thermal effects in the calculations. Yet, other
authors\cite{cheng01} have obtained a remarkable agreement with the
experimental heat of absorption from molecular dynamics simulations
that were based on a first-principles approach essentially identical
to the one used here. Thus, we can conclude our calculations are
reliable, at least at a qualitative level, for the prediction of the
absorption energetics.

A critical discrepancy between the experiment and our theory pertains
to the number of H$_2$ molecules per metal atom that can be absorbed
by the system. For stage-two C$_{24}$K, the experiments clearly
indicate a maximum uptake of two hydrogen molecules per
metal.\cite{watanabe71,lagrange72} Moreover, it is found that
stage-one C$_8$K, one of the systems we have simulated, does not
absorb any hydrogen unless significant pressures are applied: 100~atms
are necessary to obtain C$_8$K-2H$_2$.\cite{semenenko94} It is thus
obvious that a {\sl naive} theoretical prediction for large storage
capacities, as suggested by the binding energies in Table~\ref{tab1},
is not realized experimentally. Several causes might be at the origin
of this conflict. First of all, one should not forget that the
underlying LDA approximation is surely overestimating the H$_2$
binding energies at large hydrogen contents, and might also fail to
propertly describe the H$_2$--H$_2$ repulsion at short distances. In
addition, it should be noted that our theoretical representation of
the physisorbed H$_2$ does not reflect the actual {\sl effective size}
of the molecule, which is determined by the quantum nature of hydrogen
as well as by thermal effects. Indeed, in the discussion given in
Ref.~\onlinecite{watanabe73}, the effective diameter of H$_2$ is taken
to be of about 2.4~\AA; when the effective size of the metal ions is
also considered, one quickly runs into steric restrictions for the
absorption of hydrogen into the system. In particular, it is found
that there is not sufficient space for any hydrogen molecule in the
C$_8$K structure or for a third molecule, per metal atom, in the
C$_{24}$K compound. In consequence, one has to be cautious when
drawing conclusions from our simulations of hyrogen-loaded graphites,
which are characterized by very densely packed molecules. (Note that,
in the fully hydrogen-loaded lowest-energy structures, the average
H$_2$--H$_2$ distance coincides with the distance between hexagon
centers in graphene, which is about 2.46~\AA.) Indeed, a full
statistical and quantum-mechanical study would be necessary to
determine the actual significance of the lowest-energy structures
found. In principle, two scenarios are possible: such structures might
turn out to be either (i) metastable configurations that would be
accessible upon overcoming an energy barrier, or (ii) unstable
configurations that can be realized only by the continued application
of high hydrogen pressures. Given the relatively small binding enegies
per molecule that we obtained, the second possibility seems more
likely.

In conclusion, our results for graphites intercalated with alkali
metals are essentially in agreement with what was known about such
systems. Thus, they have not revealed any novel physical or chemical
effect that might have an impact in hydrogen-storage applications.

%%%

\subsection{Alkaline earth metals}

\subsubsection{Simulation results}

\begin{table}[bottom!]
\caption{Calculated binding energies (in eV) of hydrogen molecules
absorbed in stage-one graphites intercalated with alkaline earth metal
atoms. The information is organized as in
Table~\protect\ref{tab1}. The values between parenthesis correspond to
situations in which the H$_2$ molecules are physisorbed (see text). In
the ``$\bar{E}_b$'' and ``wt.\%'' columns, the values between
parenthesis are computed including the physisorption cases.}
\label{tab2}
\vskip 2mm
\begin{tabular}{lcccc|c|c}
\hline\hline
System            & n=1     & n=2    & n=3    & n=4     & $\bar{E}_b$  & wt.\% \\
\hline
C$_4$Be-nH$_2$    & 0.76    & 0.50   & (0.15) & (0.17)    & 0.63 (0.39)      & 6.6 (12.4)    \\
C$_6$Be-nH$_2$    & 0.29    & 0.70   & 0.52   & (0.05)      & 0.50 (0.39)      & 7.0  (9.1)    \\
\hline
C$_4$Mg-nH$_2$    & 0.73    & 0.50   & 0.36   & (0.08)      & 0.53 (0.41)      & 7.7 (10.0)    \\
C$_6$Mg-nH$_2$    & 0.24    & 0.25   & 0.16   & 0.24        & 0.22        &  7.7    \\
\hline
C$_4$Ca-nH$_2$    & 0.43    & $-$0.15 & 0.24   & 0.02    &  0.14   &  8.4    \\
C$_6$Ca-nH$_2$    & 0.41    & $-$0.24 & 0.34   & $-$0.02 &  0.12   &  6.7    \\
\hline\hline
\end{tabular}
\end{table}

\begin{figure}[top!]
\includegraphics[width=\columnwidth]{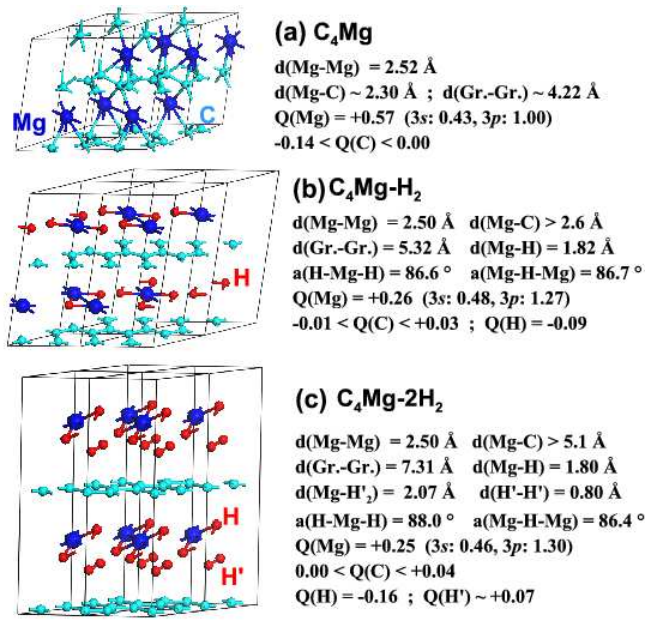}
\caption{(Color online) Lateral view of the lowest-energy
configurations obtained for the C$_4$Mg (a), C$_4$Mg-H$_2$ (b), and
C$_4$Mg-2H$_2$ (c) compositions. A 2$\times$2$\times$1 repetition of
the considered supercell is shown. The relevant distances, angles and
L\"owdin charges (in elementary charge units) are given. The in-plane
lattice parameters are essentially unaffected by the insertion of
metal and hydrogen atoms.}
\label{fig3}
\end{figure}

Most importantly, and at variance from what we obtained for the alkali
metals, in this case we find that the H$_2$ molecules react chemically
with the metal atom. The computed binding energies are given in
Table~II. The most interesting results correspond to the materials
with Be and Mg, which present a similar phenomenology and can be
described jointly.

Figure~\ref{fig3} displays the results for the C$_4$Mg-nH$_2$ systems,
which are representative of the rest. (Note that the metal density of
the C$_4$M systems is so high that the metal atoms fall within bonding
distance of each other.) As shown in Fig.~\ref{fig3}(a), the Be/Mg
atoms do not place themselves at the central point between two
eclipsed C$_6$ hexagons, but adopt a less symmetric and less
coordinated configuration. Figure~\ref{fig3}(b) shows the
lowest-energy solution resulting from the absorption of a first
H$_2$. We found that, in all the cases considered, this first hydrogen
molecule dissociates and the individual H atoms form chemical bonds
with the metal. The results for the second H$_2$ molecule are
illustrated in Fig.~\ref{fig3}(c): it places itself in the
neigbourhood of the metal atom and undergoes a considerable
elongation, the H--H bond distance increasing up to typical values
between 0.80 and 0.83~\AA. The behavior of additional H$_2$ molecules
varies from system to system, in a way that seems to be strongly
correlated with how much room there is available in the neighbourhood
of the metal atom. For example, in the case of C$_6$Be-3H$_2$ the
third molecule binds to the Be atom in the same way as the second one;
in contrast, in the more densely-packed C$_4$Be-3H$_2$ system the
third hydrogen cannot access the metal and {\sl physisorbes} into the
structure, the computed H--H distance being 0.77~\AA\ (the
corresponding binding energy is given between parenthesis in
Table~\ref{tab2}).  In both the Be and Mg cases, we found it
impossible to insert a fifth H$_2$ into the system.

The behavior of the Ca compounds is more complex. As in the cases of
Be and Mg, the first hydrogen molecule dissociates, and the individual
hydrogens form bonds with the metal. However, we found that the
addition of a second H$_2$ is not energetically favorable. Moreover,
the lowest-energy configuration is one in which there is no
dissociated hydrogen molecule. Instead, the two molecules per metal
are bound to the Ca atom and significantly elongated, the typical H--H
distance being 0.86~\AA. It is noteworthy how the dependence of the
interlayer distance with the hydrogen content reflects these binding
trends; for the C$_6$Ca-nH$_2$ systems we obtained: 4.51~\AA\ for n=0
(i.e. the system without hydrogen), 5.09~\AA\ for n=1, and 4.66~\AA\
for n=2. The relatively large interlayer spacing for the n=1 case
reflects the strong Ca--H interaction and the concomitant weakening of
the C--Ca bonds. When, for n=2, we switch to a weaker Ca--H$_2$
interaction, the interlayer distance reduces to a value that is close
to that of the n=0 case. Interestingly, the addition of a third H$_2$
molecule is energetically favorable and, in this case, the
lowest-energy solution displays two molecules weakly bound to the
metal atom, with a typical H--H distance of 0.80~\AA, and two
individual hydrogen atoms that form chemical bonds with Ca. The
interlayer distance for n=3 is computed to be 6.91~\AA, which again
reflects that the C--Ca interactions are weakened by the formation of
Ca--H bonds. A fourth H$_2$ remains molecular and its net interaction
with the rest of the system is neglegible.

\begin{figure}[top!]
\includegraphics[width=\columnwidth]{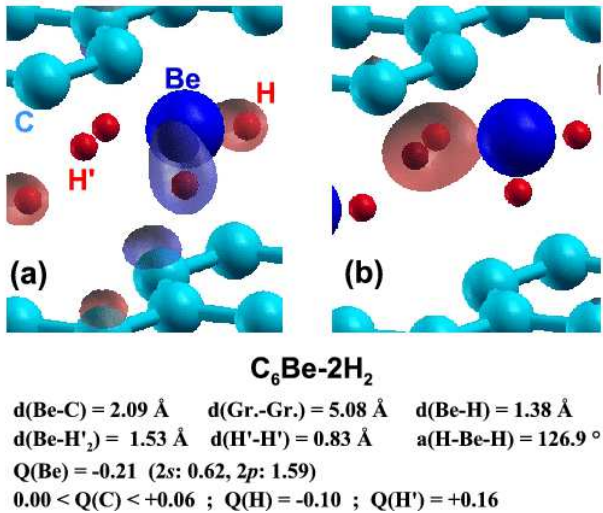}
\caption{(Color online) Computed metal--hydrogen binding eigenstates
  of the C$_6$Be-2H$_2$ system, corresponding to the $\Gamma$-point of
  the Brillouin zone.\protect\cite{xcrysden} Panel~(a): Highest
  occupied electronic state, which is associated to the bonding
  between Be and the individual H atoms. The character of the wave
  function is predominantly H-1$s$ and Be-2$p$, as evidenced by the
  different sign (different color in the figure) of the {\it lobes}
  stretching from the metal to the hydrogens. Panel~(b): Low-lying
  electronic state with predominant H-1$s$ character. In essence, this
  is the bonding $\sigma$-orbital of the H$_2$ molecule, which is only
  partly occupied as charge is transferred to Be-2$p$ orbitals. The
  relevant distances, angles, and L\"owdin charges (in elementary
  charge units) for the C$_6$Be-2H$_2$ system are given.}
\label{fig4}
\end{figure}

Also given in Figure~\ref{fig3} are the L\"owdin charges computed for
the C$_4$Mg-nH$_2$ systems. They display the following features that
are, in essence, common to all the systems studied: (1) In the
materials without hydrogen, there is a significant charge transfer
from the metal to the graphene layers; also, the empty $p$ orbitals of
the metal get populated and play an important role in the C--M
bonds. (2) The dissociation of H$_2$ is related to a considerable
charge transfer from the metal to the hydrogen atoms. As illustrated
in Fig.~\ref{fig4}(a) for the case of C$_6$Be-2H$_2$, the individual H
atoms form bonds with the metal and are negatively charged by about
0.10~electrons. (3) The interaction between the metal atoms and the
elongated H$_2$ molecules involves a charge transfer that has a marked
one-directional character: the empty $p$ orbitals of the metal drag
about 0.1-0.2~electrons from the binding $\sigma$-orbital of the
molecule (a typical low-lying M--H$_2$ binding state is displayed in
Fig.~\ref{fig4}(b)). The calculations clearly show there is no charge
back-donation from the metal to the anti-bonding $\sigma^*$-orbital of
the hydrogen molecule. (4) As the hydrogen content increases, the
graphene layers tend to donate charge to the metal--hydrogen complex.

Since our simulations predict chemical reactions and bond formation in
the case of the alkaline earth metals, it seems mandatory to ratify
the LDA results of Table~\ref{tab2} with GGA simulations, which are
well-known to be more accurate in this context.\cite{perdew96} Hence,
we performed GGA calculations for the C$_4$M-nH$_2$ systems with
M\,=\,Be and Mg. In the Be case, we obtained binding energies of
0.62~eV for n=1 and 0.31~eV for n=2; these results imply a 0.1-0.2~eV
correction (reduction) of the values in Table~\ref{tab2}. On the other
hand, for the physisorbed molecules corresponding to n=3 and n=4, the
GGA renders repulsive/neglegible interactions of $-$0.12 and 0.00~eV,
respectively. In the Mg case, the GGA binding energies for n=1, 2, and
3 are, respectively, 0.55, 0.41, and 0.03~eV. The only significant
deviation from the results in Table~\ref{tab2} pertains to the third
H$_2$ molecule, whose LDA binding energy is 0.36~eV. Such a
discrepancy probably reflects the complex nature of the binding
interactions associated to that hydrogen, which might be a combination
of chemical bonding to the metal and physisorption to the carbon
structure. Finally, the GGA binding energy for the fourth molecule,
which is undoubtedly physisorbed, is negative: $-$0.07~eV. In
conclusion, our selected GGA calculations ratify the conclusions drawn
from the LDA simulations, and indicate an LDA overestimation of about
0.1~eV for the energies associated to chemical bonds.

\subsubsection{Discussion}

We have thus found that the intercalation of alkaline earth metals
results in hydrogen-binding properties that differ significantly from
those obtained by intercalating alkali metals. The alkaline earth
metals considered react chemically with the hydrogen molecules, which
results in either (i) the dissociation of the molecule to form M--H
bonds, or (ii) the formation of a M--H$_2$ bond via a mechanism that
involves a significant charge transfer from the hydrogen molecule to
the metal. Interestingly, this difference in behavior between the
elements of the first two columns of the Periodic Table had already
been pointed out in the literature. For example,
Ref.~\onlinecite{lochan06} reports a theoretical study of the
interactions of H$_2$ with a number of free-standing ions
(e.g. Li$^{+}$, Na$^{+}$, and Mg$^{2+}$) and concludes that charge
donation from the $\sigma$ orbital of hydrogen to the low-lying empty
orbitals of the metal does occur for Mg$^{2+}$ but not for the alkali
metals; such an observation is consistent with our results. In
addition, the authors of Ref.~\onlinecite{lochan06} did not observe
any back-donation from the filled metal orbitals to the anti-bonding
$\sigma^*$-orbital of the hydrogen molecule, again in accordance with
our findings and in contrast with what is typical in
transition-metal--H$_2$ {\sl Kubas} complexes.\cite{kubas01}

Our results thus suggest that inserting alkaline earth metal atoms in
graphite-like structures should result in a significant enhancement of
the hydrogen sorption and storage properties. One might question the
reliability of this prediction, as in the case of the alkali metals
everything indicates that LDA-related errors, as well as neglected
quantum and thermal effects, add up to cause a large discrepancy
between the predicted hydrogen-storage capacity and the observations
for K-intercalated graphites. The situation seems qualitatively
different for the alkaline earths, though. First of all, in this case
the hydrogen absorption relies on the formation of chemical bonds with
the metal atoms, a situation for which DFT is known to be reasonably
accuarate. Secondly, the presence of those chemical bonds suggests
that, at variance with what is assumed for the H$_2$ molecules
physisorbed within the alkali-intercalated grahites,\cite{watanabe73}
here it does not make sense to associate a large effective diameter to
the absorbed hydrogen molecules. This is further supported by the
results of vibrational calculations, restricted to the hydrogen atoms,
that we performed for representative cases: For C$_6$Mg-2H$_2$ the
lowest-lying frequencies computed are about 270~cm$^{-1}$, reflecting
relatively strong metal--hydrogen interactions, while for
C$_{14}$Na-6H$_2$ and C$_{14}$K-6H$_2$ we obtained modes with
frequencies as low as 150~cm$^{-1}$, indicating that the hydrogen
molecules are not bound as strongly. For these reasons, we believe the
theoretical storage-capacity values reported in Table~\ref{tab2} for
the alkaline earths can be taken as relevant for the actual materials.

In summary, we can conclude that the intercalation of alkaline earth
metal atoms in graphitic carbons may result in useful materials for
hydrogen storage. It would thus be interesting to study the sorption
of hydrogen by, for example, the recently obtained stage-one C$_6$Ca
intercalated graphites.\cite{emery05a,emery05b} We hope our work will
also estimulate efforts to produce similar intercalated graphites
using Mg and Be, or to investigate carbons of the type used as
electrodes in Li batteries, but intercalated with alkaline earth
metals rather than with lithium.

\subsection{Transition metals: results and discussion}

We only studied in detail a few transition metals (Sc, Ti, V, Y, and
Zr), which was sufficient to identify general trends. Our main
findings are summarized in the following points. (All the numerical
results correspond to C$_6$M-nH$_2$ compositions.) (1) The first H$_2$
molecule interacts with the metal atom, drawing charge from it, and
dissociates. The associated binding energy increases strongly as we
move down and to the left in the Periodic Table. More precisely, we
obtained positive bindings for Sc (0.08~eV), Y (1.11~eV), and Zr
(0.23~eV), and repulsive interactions of for Ti ($-$0.30~eV) and V
($-$1.04~eV). (2) We tried the addition of a second H$_2$ molecule in
the Sc, Y, and Zr compounds. The absorption was found to be
energetically favorable only in the Sc compound, with a small
associated binding energy of 0.03~eV. The strongest repulsion
corresponds to the material with Y ($-$0.47~eV). In all the cases
considered, the second H$_2$ was found to remain molecular and bind to
the metal atom, the typical H--H bond distance obtained being about
0.86~\AA. The resulting metal--hydrogen complexes have the form
M-2H-H$_2$, as we did not observe any recombination of the H atoms
coming from the dissociation of the first hydrogen molecule. Given
that these results are not encouraging from the hydrogen-storage
perspective, as it is clear that the only hopes (if any) are
associated to relatively heavy elements, we did not consider the
insertion of a third H$_2$ molcule.

The results for the first H$_2$ molecule can be understood by noting
the strong correlation between the computed binding energies and the
graphene--graphene interlayer distances in the materials without
hydrogen. Indeed, the hydrogen-free systems display very compact
C$_6$--M--C$_6$ structures and strikingly short interlayer
distances. More precisely, for the 3$d$ metals consired, i.e. Sc, Ti,
and V, we obtained 3.82, 3.50, and 3.26~\AA, respectively; for the
4$d$ metals, Y and Zr, we got 4.30 and 4.03~\AA, respectively. (Keep
in mind that our LDA equilibrium interlayer distance for graphite is
about 3.4~\AA.) The insertion of H$_2$ causes an expansion of the
interlayer space and, naturally, such an expansion will be greater,
and energetically more costly, in the materials presenting smaller
interlayer distances in the hydrogen-free case. Indeed, we obtained
that the insertion of H$_2$ is most energetically unfavorable in the
Ti and V compounds, where the graphene--graphene distance grows up to
4.29 and 3.98~\AA, respectively, upon the aborption of H$_2$. In
contrast, it is comparatively {\sl easy} for the H$_2$ molecules to
enter the interlayer space in the Y compound (the interlayer distance
obtained in the C$_6$Y-H$_2$ case is 4.83~\AA).

In what regards the second H$_2$ molecule, the situation becomes more
complicated. Most noticeably, the Y compound, which is the material in
which more interlayer space is available, is also the one displaying a
stronger repulsion. Such a result can probably be atributed to the
existence of very stable C$_6$--(Zr-2H)--C$_6$ groups that do not gain
any energy by binding an additional H$_2$.

From our L\"owdin analysis, we did not observe any remarkable feature
concerning the C--M interactions leading to the very short interlayer
distances mentioned above. The C--M bonding seems very similar to that
occurring between transition-metal atoms and carbon nanotubes or
fullerenes, which has already been discussed in the literature (see
Ref.~\onlinecite{yil05b} and references therein). One could
nevertheless be suspicious about the LDA for this type of calculation,
since this approximation is known to overestimate the strength of
chemical bonds, which might be resulting in unrealistically small
interlayer distances. Hence, for a number of representative cases we
repeated our calculations using the GGA. For example, for the
hydrogen-free graphites with Sc, Ti, and Y, the GGA renders,
respectively, the following interlayer distances: 3.89, 3.58, and
4.41~\AA. For the same systems, the GGA binding energy for the first
H$_2$ molecule is, respectively, $-$0.18, $-$0.91, and 0.82~eV. Thus,
as we found in the case of the alkaline earth metals, the GGA corrects
the overbinding associated to the LDA, and can even switch the sign of
the interaction in cases where the LDA predicts very weak binding
(e.g. in C$_6$Sc-H$_2$). Yet, generally speaking, the GGA corrections
are not qualitatively significant and our LDA-based conclusions remain
valid.

Consequently, none of the transition metals considered appears as a
good option to enhance the hydrogen-storage capacity of graphitic
materials. Our simulations predict strong carbon--metal interactions
that preclude the possibility that the metal atoms act as convenient
binding sites for hydrogen.

\subsection{Stability of the metal-intercalated systems}

Finally, we report results concerning one of the basic motivations to
study this class of materials: the notion that graphitic systems may
bind individual metal atoms tightly enough to preclude
metal-segregation and clustering problems that would complicate
hydrogen-storage applications.

In principle, to investigate the segregation of metal atoms in these
materials one would need to perform molecular dynamics simulations
that capture the thermally activated diffusion
mechanisms. Unfortunately, that constitutes a very computationally
demanding work that falls beyond the scope of the present
study. Nevertheless, one can try to quantify the tendency to cluster
of the different elements by comparing their binding energy to the
graphitic materials with the binding energy of the corresponding
crystals. This amounts to computing the binding energy of the metal
atoms inserted in graphite {\sl taking the result for the crystalline
metal as the reference of energy}. For example, the binding energy of
Be in C$_6$Be would be:
$E$(Be~crystal)\,$+$\,$E$(graphite)\,$-$\,$E$(C$_6$Be), where
$E$(Be~crystal) is the energy per atom of the Be crystal, etc. A
positive value of a binding energy thus computed would be a strong
indication that the metal atoms will {\sl not} segregate to form
clusters. Note that, since the cohesive energy per atom is larger for
the crystal than for a cluster, the proposed criterion will tend to
overestimate the tendency to segregate. (Typically, the energy of the
isolated metal atom should be used as a reference for the computation
of the binding energy; note that binding energies thus obtained --
e.g., those reported in Ref.~\onlinecite{lugosolis07}, which discusses
the role of long-range forces in this type of systems -- cannot be
compared with ours directly.)

An important difficulty with the application of the above criterion
relates to the disparity of the forces that dominate the binding in
each of the the systems whose energies are to be compared. Indeed,
while both the LDA and GGA can be expected to be accurate for the
metallic crystals, only the LDA looks like a reasonable option for
graphite, and the GGA should be the more accurate for the
metal-intercalated materials. Being aware of these difficulties, we
have performed both LDA and GGA calculations, which allows us to have
some confidence in the trends observed. (For the GGA calculation of
$E$(graphite), we used the lowest-energy structure obtained from our
LDA calculation, as the GGA-relaxed structure deviates significantly
from the experimental one.) 

\begin{table}[bottom!]
\caption{Calculated binding energies (in eV) of metal atoms to an
  isolated graphene layer (the simulated Gr.--M systems correspond to
  the C$_{32}$M formula unit) and graphite (corresponding to the
  composition C$_6$M). Both LDA and GGA results are given. (See text
  for more details.)}
\label{tab3}
\vskip 2mm
\begin{tabular*}{\columnwidth}{@{\extracolsep{\fill}}lrrrrrrr}
\hline\hline
\multicolumn{7}{c}{LDA binding energies}\\
         &   Li      &   Na   &   K        &   Be   &  Mg   &  Ca  \\
Gr.--M   &  $-$0.40  & $-$0.56  & $-$0.16  &   $-$3.91 & $-$1.62 &  $-$1.54 \\
C$_6$M   &    0.60   & 0.20     & 0.70     &   $-$1.98  & $-$1.15 & 0.16  \\
         &  Sc       &  Ti       &  V        &  Y       &  Zr  \\
Gr.--M   &  $-$2.75  &  $-$3.89  &  $-$4.76  & $-$2.67  &  $-$4.44  & \\
C$_6$M   &  0.52  &  $-$0.46  &  $-$1.40  & 0.58     &  $-$0.43 \\
\hline
\multicolumn{7}{c}{GGA binding energies}\\
         &   Li      &   Na   &   K        &   Be   &  Mg   &  Ca  \\
Gr.--M   &  $-$0.76  & $-$0.85  & $-$0.18  &   $-$3.70 & $-$1.48 &  $-$1.54 \\
C$_6$M   &    0.38   & 0.15     & 0.44     &   $-$2.27  & $-$1.28 & 0.46  \\
         &  Sc       &  Ti       &  V        &  Y       &  Zr  \\
Gr.--M   &  $-$2.97  &  $-$3.83  &  $-$4.40  & $-$2.92  &  $-$4.42  & \\
C$_6$M   &  $-$0.03  &  $-$1.30  &  $-$1.92  & $-$0.22     &  $-$0.93
\\
\hline\hline
\end{tabular*}
\end{table}

Table~\ref{tab3} thus reports our results for the binding energies of
metal atoms to the C$_6$M systems and, for the sake of comparison, to
an isolated graphene layer. The main conclusion is that, as expected,
the metal atoms inserted in the graphitic structures can be expected
to be substantially more resistant to clustering than those deposited
on graphene. Indeed, for many elements, like the considered alkali
metals and Ca, our prediction is that the metal atoms will not
segregate at all. This result is compatible with the experimental fact
that there exist low-stage metal-intercalated carbons for Li, K, and
Ca. In what regards the alkaline earth species that are most promising
for hydrogen-storage purposes, i.e. Be and Mg, the calculations
suggest they might present a tendency towards segregation even in the
intercalated graphitic materials. The causes for the computed,
relatively weak, binding of Be and Mg to the carbon structure are
probably related with (i) the stable closed-shell electronic structure
of these atoms and (ii) their preference for coordinations that are
lower than those typical of the alkali and Ca metals (note that Be and
Mg crystalize in an hexagonal structure in which atoms are less
coordinated than in the cubic alkali and Ca metals). As for the
transition metals, the calculations also suggest a tendency to
segregate, although not as strong as for Be and Mg. Finally, let us
note that in a few cases we computed the binding energies for
different C--M ratios; the observed variations, of up to a few tenths
of eV when moving from C$_4$M to C$_{14}$M, are not significant given
the approximations involved in our estimates.

In conclusion, our results indicate that Be and Mg, the most promising
elements from the hydrogen-storage perspective, are also the most
likely ones to undergo segregation and cluster formation when
intercalated in graphitic structures. Nevertheless, given the
crudeness of the approximations made, we think this result should not
discourage experimental attempts at verifying the hydrogen-storage
mechanism predicted to occur in the Be- and Mg-intercalated materials.

\section{Summary and conclusions}

In summary, we have studied the hydrogen-binding properties of various
stage-one metal-intercalated graphites, which are idealized model
systems that we assume capture the relevant metal--hydrogen
interactions that would be present in other, structurally more
complex, graphitic materials.

In agreement with the experimental literature, we find that alkali
metals induce an expansion of the graphene--graphene interlayer
distance which, in turn, facilitates the hydrogen absorption. Hydrogen
is absorbed in molecular form and weakly bound to the carbon
structure, which results in a relatively large effective size and,
consequently, a low storage capacity. We have also studied a number of
transition metal atoms (namely, light 3$d$ and 4$d$ elements) and
obtained discouraging results in what regards hydrogen storage: our
simulations predict relatively strong carbon--metal interactions that
result in narrow interlayer spaces and prevent the metal atoms from
acting as convenient binging sites for hydrogen.

Remarkably, we obtain very promising results for the alkaline earth
metals. For all the elements studied (namely, Be, Mg, and Ca), we
observe that the intercalated metal atoms interact strongly with the
H$_2$ molecules and form metal-hydrogen complexes. Our results
indicate that each metal atom can bind a maximum of three or four
hydrogen molecules, the binding energies ranging from 0.2 to 0.7~eV,
as required for hydrogen-storage applications. The storage capacity,
predicted for the simulated stage-one graphites intercalated with Be
and Mg, can reach values in the 7-8~wt\% range.

In conclusion, our simulations indicate that alkaline earth metals are
the best option to try to obtain graphitic materials with an enhanced
hydrogen-storage performance. Some of our predictions could be
directly tested by measuring the hydrogen-absorption properties of the
recently obtained stage-one intercalated graphite
C$_6$Ca.\cite{emery05a} Yet, it must be stressed that the theory
suggests that Be and Mg based compounds may offer a superior
performance. We hope that the novel possibilities discussed in this
paper will aid current, and encourage new, experimental efforts aimed
at producing a competitive carbon-based hydrogen-storage system.

We acknowledge useful discussions with C.~Ahn, E.~Canadell,
M.H.~Cohen, C.~Contescu, K.S.~Nahm, G.~Walker, and T.~Yildirim. This
is work made in the context of our participation in Task~22 of the
International Energy Agency's Hydrogen Implementing Agreement. It was
finacially supported by the Spanish Ministry of Science and Education
(ENE2006-27257-E, FIS2006-12117-C04-01 and CSD2007-00041) and the
Catalan Government (SGR2005-683).

\end{document}